# Kinetic Alfvén waves in the temperature anisotropic space plasma with a kappa-Maxwellian distribution

Rui Huo[1], Jiulin Du[1] and Ran Guo[2]

1. *Department of Physics, School of Science, Tianjin University, Tianjin 300350, China*
2. *Department of Physics, College of Science, Civil Aviation University of China, Tianjin 300300, China*

**Abstract** The dispersion and damping rate of kinetic Alfvén waves are studied in temperature anisotropic space plasma with kappa-Maxwellian distribution. Employing a kinetic approach, the wave frequency and damping rate of kinetic Alfvén waves and the modified ion acoustic waves are derived in a low *β* plasma, which both depend on the parameters $\kappa$ and $T_{\perp e}/T_e$. The numerical analyses show that the wave frequency $\omega_r/kV_A$ of kinetic Alfvén waves is larger in kappa-Maxwellian plasma than that in Maxwellian case. The wave frequency of the modified ion acoustic waves in kappa-Maxwellian plasma is larger in the short-wave region but smaller in the long-wave region than that in Maxwellian case. Again, we found that the damping rate of kinetic Alfvén waves in kappa-Maxwellian plasma is stronger than that in Maxwellian case. The damping rate of modified ion acoustic waves in kappa-Maxwellian plasma is stronger in the short-wave region but weaker in the long-wave region than that in Maxwellian case. The impact of the parameter $T_{\perp e}/T_e$ on the two modes is relatively small because we consider the low *β* case. These results are helpful for us to understand better the characteristics of kinetic Alfvén waves in space plasma.

**Keywords:** Kinetic Alfvén waves; Modified ion acoustic waves; Dispersion relation; Kappa-Maxwellian distribution

## 1. Introduction

Alfvén waves are the basic low-frequency and long-wavelength wave modes in magneto-plasmas. It was first predicted by Alfvén in 1942 [1]. Kinetic Alfvén waves are dispersive Alfvén waves with a short perpendicular wavelength comparable to ion gyroradius or the electron inertial length. In this case, ions will not follow the magnetic field lines, but the electrons still do because of the smallness of their Larmor radius. This causes a charge separation and generates the kinetic Alfvén waves [2]. Kinetic Alfvén waves have become an increasingly interesting and extensively studied subject because they can play important roles in particle energization and energy transport in space, and solar and other astrophysical plasma [3, 4].

Hasegawa made a significant contribution to the kinetic Alfvén waves [5, 6]. He proposed that kinetic Alfvén waves can transfer wave energy to electrons through Landau damping, thus leading to plasma heating or electron acceleration along the magnetic field. Lysak and Lotko presented the kinetic Alfvén waves in space plasma with a Maxwellian distribution [7]. They mainly considered the impact of full ion and electron gyroradius and full electron kinetics on the kinetic Alfvén waves in the low frequency and low *β* plasma, where *β* is a ratio of the thermal pressure to the magnetic pressure. On this basis, Bashir et al. studied the influence of temperature anisotropy on the waves [8]. Subsequently, Barik et al. studied the generation of kinetic Alfvén waves by ion beam and velocity shear in the Earth's magnetosphere [9].

We note that the researches above are studied in the case of plasmas with a Maxwellian distribution. However, most particles in space plasmas are observed to be out of thermal equilibrium. The particle velocity distributions observed display the suprathermal tails [10] and temperature anisotropies [11]. The appearance of energetic particles causes the plasma to deviate from thermal equilibrium, therefore the associated velocity distributions are non-Maxwell ones. In fact, non-Maxwellian distributions are very common in nonequilibrium astrophysical and space plasmas, such as kappa distribution [12], *q*-distribution in nonextensive statistics [13], Cairns-Tsallis distribution [14], and so on. By employing a kappa distribution for space plasma, Gaelzer and Ziebell have investigated the dispersion and damping of dispersive Alfvén waves [15]. Liu et al. investigated the dispersion and damping of dispersive Alfvén waves in the plasma under the framework of nonextensive statistics [16]. The particle velocity distributions considering the temperature anisotropy are also widely used in the theoretical study of dispersive Alfvén wave [17, 18]. In some physical circumstances, one may expect that the distribution function takes a Maxwellian form in a plane perpendicular to the background magnetic field, but it takes a power-law form (e.g., the kappa distribution) in the parallel direction because of acceleration in the direction of the magnetic field. In this consideration, a kappa-Maxwell velocity distribution was introduced, expressed for the α components (electrons or ions) [19] as

$$f_\alpha(v_\parallel, v_\perp) = \frac{1}{\pi^{3/2} v_{T\perp\alpha}^2 v_{T\parallel\alpha}} \frac{\Gamma(\kappa+1)}{\kappa^{3/2}\Gamma\left(\kappa-\frac{1}{2}\right)} \times \left(1 + \frac{v_\parallel^2}{\kappa v_{T\parallel\alpha}^2}\right)^{-\kappa} \exp\left(-\frac{v_\perp^2}{v_{T\perp\alpha}^2}\right), \quad (1)$$

where $\kappa > 3/2$ is a parameter and if $\kappa \to \infty$ the kappa-Maxwell distribution (1) becomes the bi-Maxwellian distribution, the parallel and perpendicular thermal velocity are, respectively,

$$v_{T\parallel\alpha} = \sqrt{\frac{\kappa - 3/2}{\kappa} \frac{2T_{\parallel\alpha}}{m_\alpha}}, \text{ and } v_{T\perp\alpha} = \sqrt{2T_{\perp\alpha}/m_\alpha}.$$

The kappa-Maxwellian distribution was observed by FREJA and Voyager PLS satellites in the space plasmas, such as the auroral region and Saturn's magnetosphere [20, 21]. It has been widely applied to describe various plasma phenomena and processes. In theory, Basu studied the linear dispersion relations of electrostatic waves in the spatially inhomogeneous anisotropic plasma with kappa-Maxwellian distribution [22]. Nazeer et al. discussed the propagation of electromagnetic electron cyclotron waves by employing kappa-Maxwellian distribution for energetic trapped electrons in the auroral region [23]. They also investigated the electron firehose instability in the space plasmas with kappa-Maxwellian distribution [24]. In this work, we employ the kappa-Maxwellian distribution to study the dispersion and damping of the kinetic Alfvén waves.

The paper is organized as follows. In section 2, the dispersion relation and the damping rate of kinetic Alfvén waves are derived from the kinetic theory. In section 3, numerical analyses are made. In section 4, the conclusion is given.

## 2. The dispersion and damping of kinetic Alfvén waves

We now introduce the basic theory of kinetic Alfvén waves in the collisionless magnetized electron-ion plasma. The kinetic model is described by a set of Vlasov-Maxwell equations,

$$\left[\frac{\partial}{\partial t}+\mathbf{v}\cdot\frac{\partial}{\partial \mathbf{r}}+\frac{q_\alpha}{m_\alpha}(\mathbf{E}+\frac{\mathbf{v}\times\mathbf{B}}{c})\cdot\frac{\partial}{\partial \mathbf{v}}\right]f_\alpha(\mathbf{r},\mathbf{v},t)=0, \tag{2}$$

$$\nabla\times\mathbf{E}=-\frac{1}{c}\frac{\partial \mathbf{B}}{\partial t}, \tag{3}$$

$$\nabla\times\mathbf{B}=\frac{4\pi}{c}\mathbf{J}+\frac{1}{c}\frac{\partial \mathbf{E}}{\partial t}, \tag{4}$$

where $f_\alpha(\mathbf{r}, \mathbf{v}, t)$ is the velocity distribution function of particles. $\mathbf{E}$ and $\mathbf{B}$ are the electric field intensity and the magnetic induction, respectively; $q_\alpha$ and $m_\alpha$ are respectively the charge and mass of the particle, and $c$ is the light speed. Let the velocity distribution function be a stationary-state distribution $f_{\alpha 0}$ and a small disturbance $f_{\alpha 1}$, namely, $f_\alpha(\mathbf{r}, \mathbf{v}, t) = f_{\alpha 0}(\mathbf{r}, \mathbf{v}) + f_{\alpha 1}(\mathbf{r}, \mathbf{v}, t)$, also let the electric field intensity $\mathbf{E}(\mathbf{r}, t)= \mathbf{E_0}(\mathbf{r}) + \mathbf{E_1}(\mathbf{r}, t)$ and the magnetic induction $\mathbf{B}(\mathbf{r}, t) = \mathbf{B_0}(\mathbf{r}) + \mathbf{B_1}(\mathbf{r}, t)$, where $\mathbf{E_1}(\mathbf{r}, t)$ and $\mathbf{B_1}(\mathbf{r}, t)$ are small perturbations. Taking the perturbations in the form of $f_{\alpha 1}$, $\mathbf{E_1}$, $\mathbf{B_1} \propto \exp[i(\mathbf{k}\cdot\mathbf{r}-\omega t)]$, by solving the above equations, then we can have

$$i(\omega-\mathbf{k}\cdot\mathbf{v})f_{\alpha 1}-\frac{q_\alpha}{m_\alpha}\left(\frac{\mathbf{v}\times\mathbf{B_0}}{c}\right)\cdot\frac{\partial f_{\alpha 1}}{\partial \mathbf{v}}=\frac{q_\alpha}{m_\alpha}\left(\mathbf{E_1}+\frac{\mathbf{v}\times(\mathbf{k}\times\mathbf{E_1})}{\omega}\right)\cdot\frac{\partial f_{\alpha 0}}{\partial \mathbf{v}}, \tag{5}$$

where we have used $\dfrac{\mathbf{B_1}}{c}=\dfrac{\mathbf{k}\times\mathbf{E_1}}{\omega}$. We adopt the gyro-coordinates $(v_\perp,\varphi,v)$ in the $\mathbf{v}$-space with its z-axis parallel to $\mathbf{B_0}=B_0\hat{\mathbf{z}}$, so that $v_x=v_\perp\cos\varphi$, $v_y=v_\perp\sin\varphi$, $v_z=v$. Then we can obtain

$$\frac{q_\alpha}{m_\alpha c}(\mathbf{v}\times B_0)\cdot\frac{\partial f_{\alpha 1}}{\partial \mathbf{v}}=-\Omega_\alpha\frac{\partial f_{\alpha 1}}{\partial \varphi}, \tag{6}$$

where $\Omega_\alpha=q_\alpha B_0/m_\alpha c$ is the gyro-frequency of the particle. Combining Eq. (5) and (6), we can have

$$\frac{\partial f_{\alpha 1}}{\partial \varphi}=-i\frac{\omega-\mathbf{k}\cdot\mathbf{v}}{\Omega_\alpha}f_{\alpha 1}+\frac{q_\alpha}{m_\alpha}\frac{1}{\Omega_\alpha}\left(\mathbf{E_1}+\frac{\mathbf{v}\times(\mathbf{k}\times\mathbf{E_1})}{\omega}\right)\cdot\frac{\partial f_{\alpha 0}}{\partial \mathbf{v}}. \tag{7}$$

We assume that wave vector $\mathbf{k}$ is in the x-z plane, $\mathbf{k}=(k_\perp,0,k_\parallel)$. Multiplying the integrating factor $\exp\left(-\dfrac{i}{\Omega_\alpha}\int_0^\varphi(\mathbf{k}\cdot\mathbf{v}-\omega)d\varphi\right)$ on both sides of Eq. (7), and then integrating it. After some algorithms, we can get

$$f_{\alpha 1}=\frac{iq_\alpha}{m_\alpha}\exp(i\mu_\alpha\sin\varphi)\sum_{n=-\infty}^{n=\infty}\frac{\mathbf{a}_n\cdot\mathbf{E}}{k_\parallel v_\parallel-\omega+n\Omega_\alpha}\exp(-in\varphi), \tag{8}$$

where the vector $\mathbf{a}_n=(a_{n1},a_{n2},a_{n3})$ with

$$a_{n1}=\left[\frac{k v_\perp}{\omega}\cdot\frac{\partial f_{\alpha 0}}{\partial v}+\left(1-\frac{k v}{\omega}\right)\frac{\partial f_{\alpha 0}}{\partial v_\perp}\right]\frac{n}{\lambda_\alpha}J_n(\mu_\alpha),$$

$$a_{n2}=\left[\frac{k v_\perp}{\omega}\cdot\frac{\partial f_{\alpha 0}}{\partial v}+\left(1-\frac{k v}{\omega}\right)\frac{\partial f_{\alpha 0}}{\partial v_\perp}\right]iJ'_n(\mu_\alpha),$$

$$a_{n3}=\left[\left(1-\frac{n\Omega_\alpha}{\omega}\right)\frac{\partial f_{\alpha 0}}{\partial v}+\frac{n\Omega_\alpha}{\omega}\cdot\frac{v}{v_\perp}\cdot\frac{\partial f_{\alpha 0}}{\partial v_\perp}\right]J_n(\mu_\alpha),$$

where $J_n(\mu_\alpha)$ is a Bessel function with $\mu_\alpha=k_\perp v_\perp/\Omega_\alpha$. By substituting (8) into the current density $\mathbf{j}_1 = \sum_\alpha q_\alpha \int \mathbf{v} f_{\alpha 1} d\mathbf{v}$ and $\mathbf{j}_1 = \boldsymbol{\sigma} \cdot \mathbf{E}_1$, the conductivity tensor is obtained [25] as

$$\boldsymbol{\sigma} = i \sum_\alpha \frac{q_\alpha^2}{m_\alpha \omega} \sum_{n=-\infty}^{\infty} \int d\mathbf{v} \frac{\mathbf{S}_n(v_\perp, v_\parallel; n)}{\omega - k_\parallel v_\parallel - n\Omega_\alpha}, \tag{9}$$

where

$$\mathbf{S}_n(v_\perp, v_\parallel; n) = \begin{bmatrix} v_\perp U \left(\frac{nJ_n}{\mu_\alpha}\right)^2 & iv_\perp U \frac{n}{\mu_\alpha} J_n J_n' & v_\perp W \frac{n}{\mu_\alpha} J_n^2 \\ -iv_\perp U \frac{n}{\mu_\alpha} J_n J_n' & v_\perp U (J_n')^2 & -iv_\perp W J_n J_n' \\ v_\parallel U \frac{n}{\mu_\alpha} J_n^2 & iv_\parallel U J_n J_n' & v_\parallel W J_n^2 \end{bmatrix}$$

with

$$U = (\omega - k_\parallel v_\parallel) \frac{\partial f_{\alpha 0}}{\partial v_\perp} + k_\parallel v_\perp \frac{\partial f_{\alpha 0}}{\partial v_\parallel}$$

and

$$W = \frac{n\Omega_\alpha v_\parallel}{v_\perp} \frac{\partial f_{\alpha 0}}{\partial v_\perp} + (\omega - n\Omega_\alpha) \frac{\partial f_{\alpha 0}}{\partial v_\parallel}.$$

And then we obtain the dielectric tensor

$$\boldsymbol{\varepsilon} = \mathbf{I} + \frac{i}{\varepsilon_0 \omega} \boldsymbol{\sigma} = \mathbf{I} + \sum_\alpha \frac{\omega_{p\alpha}^2}{n_{\alpha 0} \omega^2} \sum_{n=-\infty}^{\infty} \int d\mathbf{v} \frac{\mathbf{S}_n(v_\perp, v_\parallel; n)}{\omega - k_\parallel v_\parallel - n\Omega_\alpha}, \tag{10}$$

where $\omega_{p\alpha} = \sqrt{4\pi n_0 q_\alpha^2 / m_\alpha}$ is the plasma frequency, and the dispersion relation for the electromagnetic waves in the plasma [25],

$$\begin{vmatrix} \varepsilon_{xx} - n_\parallel^2 & \varepsilon_{xy} & \varepsilon_{xz} + n_\parallel n_\perp \\ -\varepsilon_{xy} & \varepsilon_{yy} - n^2 & \varepsilon_{yz} \\ \varepsilon_{zx} + n_\parallel n_\perp & -\varepsilon_{yz} & \varepsilon_{zz} - n_\perp^2 \end{vmatrix} = 0, \tag{11}$$

where $n_\parallel = ck_\parallel/\omega$, $n_\perp = ck_\perp/\omega$, and $n = ck/\omega$ are the parallel, perpendicular, and total refractive indices, respectively. For the case of the plasma with the low frequency, the long parallel wavelength and the low $\beta$ (i.e., $\beta = 4\pi n_0 T_{\parallel e}/B_0^2 = c_s^2/V_A^2 \ll 1$), the dispersion relation (11) for the kinetic Alfvén waves [7] becomes

$$(\varepsilon_{xx} - n_\parallel^2)(\varepsilon_{zz} - n_\perp^2) - n_\parallel^2 n_\perp^2 = 0, \tag{12}$$

where

$$\varepsilon_{xx} = 1 + \sum_\alpha \frac{\omega_{p\alpha}^2}{\omega} \int v_\perp d^3v \sum_{n=-\infty}^{\infty} \left(\frac{n^2}{\mu^2} J_n^2(\mu_\alpha)\right) \times \frac{\left(1 - \frac{k_\parallel v_\parallel}{\omega}\right)\frac{\partial f_{\alpha 0}}{\partial v_\perp} + \frac{k_\parallel v_\perp}{\omega}\frac{\partial f_{\alpha 0}}{\partial v_\parallel}}{\omega - k_\parallel v_\parallel - n\Omega_\alpha}, \tag{13}$$

$$\varepsilon_{zz} = 1 + \sum_\alpha \frac{\omega_{p\alpha}^2}{\omega} \int v_\perp d^3v \sum_{n=-\infty}^{\infty} \left(J_n^2(\mu_\alpha) \frac{v_\parallel}{v_\perp}\right) \times \frac{\frac{n\Omega_\alpha}{\omega} \frac{v_\parallel}{v_\perp} \frac{\partial f_{\alpha 0}}{\partial v_\perp} + \left(1 - \frac{n\Omega_\alpha}{\omega}\right)\frac{\partial f_{\alpha 0}}{\partial v_\parallel}}{\omega - k_\parallel v_\parallel - n\Omega_\alpha}. \tag{14}$$

Here we consider the compensated electron-ion plasma ($\alpha = e, i$) with the kappa-Maxwellian distribution given by $f_{\alpha 0}$ in Eq. (1), for the low frequency Alfven wave, we have $\Omega_\alpha \gg \omega, k_\parallel v_{T\parallel\alpha}$ (thus $|\omega - n\Omega_\alpha|/k_\parallel v_{T\parallel\alpha} \gg 1$), and we can derive (see Appendix A) that

$$\varepsilon_{xx} = 1 + \frac{c^2}{V_A^2}\left(\frac{1-A_0(\lambda_i)}{\lambda_i}\right) - \frac{c^2 k_\parallel^2}{\omega^2}\chi_1, \tag{15}$$

$$\varepsilon_{zz} = 1 - \frac{A_0(\lambda_e)}{2k_\parallel^2 \lambda_{De}^2}\frac{d}{d\eta_e}Z_{\kappa,M}(\eta_e) - \frac{A_0(\lambda_i)}{2k_\parallel^2 \lambda_{Di}^2}\frac{d}{d\eta_i}Z_{\kappa,M}(\eta_i) + \frac{\omega_{pi}^2}{\omega^2}\chi_2, \tag{16}$$

where

$$\chi_1 = \frac{c_s^2}{V_A^2}\left(\left(\frac{1-A_0(\lambda_e)}{\lambda_e}\right)\left(\frac{\kappa-3/2}{\kappa}\frac{T_{\perp e}}{T_{\parallel e}}-1\right) + \left(\frac{1-A_0(\lambda_i)}{\lambda_i}\right)\left(\frac{\kappa-3/2}{\kappa}\frac{T_{\perp i}}{T_{\parallel i}}-1\right)\frac{T_{\parallel i}}{T_{\parallel e}}\right)$$

with $c_s = \sqrt{T_{\parallel e}/m_i}$ and $V_A = B_0/\sqrt{4\pi n_0 m_i}$. And in Eq. (16), we have denoted $\lambda_{D\alpha}^2 = v_{T\parallel\alpha}^2/2\omega_{p\alpha}^2$,

$$\chi_2 = \frac{m_i}{m_e}[1-A_0(\lambda_e)]\left(\frac{T_{\parallel e}}{T_{\perp e}}\frac{\kappa}{\kappa-3/2}-1\right) + [1-A_0(\lambda_i)]\left(\frac{T_{\parallel i}}{T_{\perp i}}\frac{\kappa}{\kappa-3/2}-1\right).$$

In the kinetic limit, we have $v_{T\,i} \ll \omega/k \ll v_{T\,e}$ and $m_e/m_i \ll \beta \ll 1$. Considering the fact that the ion mass is much larger than the electron mass, the collisionless damping caused by the ions is neglected [16]. In this case, the derivative of the kappa-Maxwellian distribution plasma function can be written as [19]

$$\frac{d}{d\eta_e}Z_{\kappa,M}(\eta_e) = -\frac{2\kappa-1}{\kappa} - 2\eta_e i\sqrt{\pi}\frac{\Gamma(\kappa+1)}{\kappa^{3/2}\Gamma(\kappa-1/2)}\left(1+\frac{\eta_e^2}{\kappa}\right)^{-\kappa},$$

$$\frac{d}{d\eta_i}Z_{\kappa,M}(\eta_i) = \frac{1}{\eta_i^2}. \tag{17}$$

Because $k^2\lambda_{De}^2 \ll 1$ and $c^2/V_A^2 \gg 1$, the term unity in Eqs. (15) and (16) can be removed. Taking the elements $\varepsilon_{xx}$ and $\varepsilon_{zz}$ into Eq. (10), the dispersion relation becomes

$$D_{\kappa,M}(\omega,k) \equiv \left[\frac{\omega^2}{k_\parallel^2 V_A^2} - \left(\frac{\lambda_i}{1-A_0(\lambda_i)}\right)(1+\chi_1)\right]\left(-\frac{A_0(\lambda_e)}{2k_\parallel^2 \lambda_{De}^2}\frac{d}{d\eta_e}Z_{\kappa,M}(\eta_e) - \frac{A_0(\lambda_i)}{2k_\parallel^2 \lambda_{Di}^2}\frac{d}{d\eta_i}Z_{\kappa,M}(\eta_i) + \frac{\omega_{pi}^2}{\omega^2}\chi_2\right)$$

$$-\left[\frac{\omega^2}{k_\parallel^2 V_A^2} - \left(\frac{\lambda_i}{1-A_0(\lambda_i)}\right)\chi_1\right]\frac{c^2 k_\perp^2}{\omega^2} = 0,$$

(18)

where $D_{\kappa,M}(\omega, k)$ is the dispersion relation of the kinetic Alfvén waves in the plasma with the kappa-Maxwell distribution. Therefore, using the above expressions, we have that

$$D_{\kappa,M}(\omega,k) = \left[\omega^2 - k_\parallel^2 V_A^2 \frac{\lambda_i(1+\chi_1)}{1-A_0(\lambda_i)}\right]\left\{\frac{\omega^2}{k_\parallel^2 \lambda_{De}^2 \omega_{pi}^2}\left[\frac{2\kappa-1}{4\kappa} + i\sqrt{\pi}\eta_e\frac{\Gamma(\kappa+1)}{\kappa^{3/2}\Gamma(\kappa-1/2)}\left(1+\frac{\eta_e^2}{\kappa}\right)^{-\kappa}\right]\right.$$

$$\left. -\frac{\omega^2 A_0(\lambda_i)}{A_0(\lambda_e)2k_\parallel^2 \lambda_{Di}^2 \omega_{pi}^2}\frac{1}{\eta_i^2} + \frac{\chi_2}{A_0(\lambda_e)}\right\} - \left(\omega^2 - \frac{\lambda_i k_\parallel^2 V_A^2}{1-A_0(\lambda_i)}\chi_1\right)\frac{c^2 k_\perp^2}{A_0(\lambda_e)\omega_{pi}^2} = 0. \tag{19}$$

For the small gyro-radii and low $\beta$ plasma [8], i.e., $c_s^2/V_A^2 \ll 1$, we have

$$A_0(\lambda_{e,i}) \approx 1 - \lambda_{e,i} + \frac{3}{4}\lambda_{e,i}^2,$$

$$\frac{\lambda_{e,i}}{1-A_0(\lambda_{e,i})} = \frac{1}{1-\frac{3}{4}\lambda_{e,i}} \simeq 1 + \frac{3}{4}\lambda_{e,i} = 1 + \frac{3}{4}k_\perp^2 \rho_{e,i}^2. \tag{20}$$

Inserting $\omega = \omega_r + i\gamma$ into Eq. (19), in the case of $\gamma \ll \omega_r$, we have that

$$D_{\kappa,M}(\omega,k) = \mathrm{Re}\, D_{\kappa,M}(\omega,k) + i\, \mathrm{Im}\, D_{\kappa,M}(\omega,k)$$
$$\approx \mathrm{Re}\, D_{\kappa,M}(\omega_r,k) + i\gamma \frac{\partial \mathrm{Re}\, D_{\kappa,M}(\omega_r,k)}{\partial \omega_r} + i\, \mathrm{Im}\, D_{\kappa,M}(\omega_r,k) = 0. \tag{21}$$

$$\mathrm{Re}\, D_{\kappa,M}(\omega_r,k) = 0, \quad \text{and} \quad \gamma = -\frac{\mathrm{Im}\, D_{\kappa M}(\omega_r,k)}{\partial \mathrm{Re}\, D_{\kappa M}(\omega_r,k)/\partial \omega_r}. \tag{22}$$

By solving Eq. (22) we can get the wave frequency and the damping rate (the growing rate) of kinetic Alfvén waves. After some calculations, we can get

$$\mathrm{Re}\, D_{\kappa,M}(\omega_r,k) = \frac{2\kappa-1}{2\kappa}\omega_r^4 - \omega_r^2 k_\parallel^2 \left\{ V_A^2 \left[ \frac{2\kappa-1}{2\kappa}\left(1+\frac{3}{4}k_\perp^2\rho_i^2\right)\left(1+\chi_1'\frac{c_s^2}{V_A^2}\right) + \frac{T_{\parallel e}}{T_{\parallel i}}k_\perp^2\rho_i^2 \right] + c_s^2\left[1 - k_\perp^2\rho_i^2(1+\chi_2')\right] \right\}$$
$$+ c_s^2 k_\parallel^4 V_A^2 \left(1+\frac{3}{4}k_\perp^2\rho_i^2\right)\left[1 - k_\perp^2\rho_i^2 \chi_2'\left(1+\chi_1'\frac{c_s^2}{V_A^2}\right) + \chi_1'\frac{c_s^2}{V_A^2}\left(1 - k_\perp^2\rho_i^2 + \frac{c^2 k_\perp^2}{\omega_{pi}^2}\right)\right], \tag{23}$$

$$\mathrm{Im}\, D_{\kappa,M}(\omega_r,k) = \frac{\omega_r^3}{k_\parallel v_{T\parallel e}}\left[\omega_r^2 - k_\parallel^2 V_A^2\left(1+\frac{3}{4}k_\perp^2\rho_i^2\right)\left(1+\frac{c_s^2}{V_A^2}\chi_1'\right)\right]\frac{\sqrt{\pi}\Gamma(\kappa+1)}{\kappa^{3/2}\Gamma(\kappa-1/2)}\left(1+\frac{\omega_r^2}{\kappa k_\parallel^2 v_{T\parallel e}^2}\right)^{-\kappa}, \tag{24}$$

where we denote

$$\chi_1' = \left(\frac{\kappa-3/2}{\kappa}\frac{T_{\perp e}}{T_{\parallel e}} - 1\right) + \left(1 - \frac{3}{4}k_\perp^2\rho_i^2\right)\left(\frac{\kappa-3/2}{\kappa}\frac{T_{\perp i}}{T_{\parallel i}} - 1\right)\frac{T_{\parallel i}}{T_{\parallel e}},$$

$$\chi_2' = \frac{T_{\perp e}}{T_{\parallel e}}\left(\frac{T_{\parallel e}}{T_{\perp e}}\frac{\kappa}{\kappa-3/2} - 1\right) + \frac{T_{\parallel i}}{T_{\perp i}}\frac{\kappa}{\kappa-3/2} - 1.$$

Making $\mathrm{Re}\, D_{\kappa,M}(\omega_r,k) = 0$, we have

$$\frac{\omega_{r1}^2}{k_\parallel^2 V_A^2} = \left(\frac{2\kappa-1}{2\kappa} + \frac{2\kappa-1}{2\kappa}\frac{3}{4}k_\perp^2\rho_i^2 + \frac{T_{\parallel e}}{T_{\parallel i}}k_\perp^2\rho_i^2 + \chi_1'\frac{2\kappa-1}{2\kappa}\left(1+\frac{3}{4}k_\perp^2\rho_i^2\right)\frac{c_s^2}{V_A^2} + \frac{c_s^2}{V_A^2}\left(1 - k_\perp^2\rho_i^2(1+\chi_2')\right)\right)$$
$$\times \left[\frac{2\kappa}{2\kappa-1} - \frac{\left(\frac{c_s^2}{V_A^2}\left(1+\left(\frac{3}{4}+\frac{T_{\parallel i}}{T_{\parallel e}}\frac{m_e}{m_i}\right)k_\perp^2\rho_i^2\right)\left(1 - k_\perp^2\rho_i^2\left(\chi_2'\left(1+\chi_1'\frac{c_s^2}{V_A^2}\right)\right) + \chi_1'\frac{c_s^2}{V_A^2}\left(1 - k_\perp^2\rho_i^2 + \frac{c^2 k_\perp^2}{\omega_{pi}^2}\right)\right)\right)}{\left(\left(\frac{2\kappa-1}{2\kappa} + \frac{2\kappa-1}{2\kappa}\frac{3}{4}k_\perp^2\rho_i^2 + \frac{T_{\parallel e}}{T_{\parallel i}}k_\perp^2\rho_i^2 + \chi_1'\frac{2\kappa-1}{2\kappa}\left(1+\frac{3}{4}k_\perp^2\rho_i^2\right)\right)\frac{c_s^2}{V_A^2} + \frac{c_s^2}{V_A^2}\left(1 - k_\perp^2\rho_i^2(1+\chi_2')\right)\right)^2}\right],$$

(25)

$$\frac{\omega_{r2}^2}{k_\parallel^2 V_A^2} = \left(\frac{2\kappa-1}{2\kappa} + \frac{2\kappa-1}{2\kappa}\frac{3}{4}k_\perp^2\rho_i^2 + \frac{T_{\parallel e}}{T_{\parallel i}}k_\perp^2\rho_i^2 + \frac{2\kappa-1}{2\kappa}\left(1+\frac{3}{4}k_\perp^2\rho_i^2\right)\frac{c_s^2}{V_A^2}\chi_1' + \frac{c_s^2}{V_A^2}\left(1 - k_\perp^2\rho_i^2(1+\chi_2')\right)\right)$$
$$\times \frac{\left(\frac{c_s^2}{V_A^2}\left(1+\left(\frac{3}{4}+\frac{T_{\parallel i}}{T_{\parallel e}}\frac{m_e}{m_i}\right)k_\perp^2\rho_i^2\right)\left(1 - k_\perp^2\rho_i^2\left(\chi_2'\left(1+\chi_1'\frac{c_s^2}{V_A^2}\right)\right) + \chi_1'\frac{c_s^2}{V_A^2}\left(1 - k_\perp^2\rho_i^2 + \frac{c^2 k_\perp^2}{\omega_{pi}^2}\right)\right)\right)}{\left(\left(\frac{2\kappa-1}{2\kappa} + \frac{2\kappa-1}{2\kappa}\frac{3}{4}k_\perp^2\rho_i^2 + \frac{T_{\parallel e}}{T_{\parallel i}}k_\perp^2\rho_i^2 + \chi_1'\frac{2\kappa-1}{2\kappa}\left(1+\frac{3}{4}k_\perp^2\rho_i^2\right)\frac{c_s^2}{V_A^2}\right) + \frac{c_s^2}{V_A^2}\left(1 - k_\perp^2\rho_i^2(1+\chi_2')\right)\right)^2}. \tag{26}$$

Eq. (25) and (26) give the wave frequency of the kinetic Alfvén waves and the modified ion acoustic waves, respectively. When we take $\kappa \to \infty$, the results of the plasma with Maxwellian distribution in Ref. [8] are retrieved. We also can get the damping rate (the growing rate) of kinetic Alfvén waves and modified ion acoustic waves, it can be expressed as

$$\frac{\gamma}{\omega_r} = -\frac{1}{\omega_r} \frac{\operatorname{Im} D_{\kappa M}(k,\omega_r)}{\partial \operatorname{Re} D_{\kappa M}(k,\omega_r)/\partial \omega_r}$$

$$= -\frac{\left[\frac{\omega_r^2}{k_\parallel^2 V_A^2} - \left(1 + \frac{3}{4} k_\perp^2 \rho_i^2\right)\left(1 + \frac{c_s^2}{V_A^2} \chi_1'\right)\right] \frac{\sqrt{\pi}\Gamma(\kappa+1)}{\kappa^{3/2}\Gamma(\kappa-1/2)} \frac{\omega_r}{k_\parallel v_{T\parallel e}} \left(1 + \frac{\omega_r^2}{\kappa k_\parallel^2 v_{T\parallel e}^2}\right)^{-\kappa}}{\frac{4\kappa-2}{\kappa} \frac{\omega_r^2}{k_\parallel^2 V_A^2} - \frac{2\kappa-1}{\kappa}\left[1 + \frac{3}{4}k_\perp^2\rho_i^2 + \chi_1'\left(1 + \frac{3}{4}k_\perp^2\rho_i^2\right)\frac{c_s^2}{V_A^2}\right] + \frac{2T_{\parallel e}}{T_{\parallel i}}k_\perp^2\rho_i^2 + \frac{2c_s^2}{V_A^2}\left[1 - k_\perp^2\rho_i^2(1+\chi_2')\right]},$$

(27)

where

$$\chi_1' = \left(\frac{\kappa-\frac{3}{2}}{\kappa} \frac{T_{\perp e}}{T_{\parallel e}} - 1\right) + \left(1 - \frac{3}{4}k_\perp^2\rho_i^2\right)\left(\frac{\kappa-\frac{3}{2}}{\kappa} \frac{T_{\perp i}}{T_{\parallel i}} - 1\right)\frac{T_{\parallel i}}{T_{\parallel e}},$$

$$\chi_2' = \frac{T_{\perp e}}{T_{\parallel e}}\left(\frac{T_{\parallel e}}{T_{\perp e}} \frac{\kappa}{\kappa-\frac{3}{2}} - 1\right) + \frac{T_{\parallel i}}{T_{\perp i}} \frac{\kappa}{\kappa-\frac{3}{2}} - 1.$$

## 3. Numerical analyses

In the previous section, we derived the analytical forms (in approximative case) of the frequency in Eqs. (25)-(26) and the damping rate in Eq. (27) of kinetic Alfvén waves and modified ion acoustic waves. Here, to be rigorous, we make numerical analyses of the dispersion relation in Eq. (18) to show the effect of the parameter $\kappa$ and electrons temperature anisotropic $T_{\perp e}/T_e$ on kinetic Alfvén waves and modified ion acoustic waves. We have used the parameters appropriate to the plasma sheet boundary layer at altitudes of 4–6 in Re [26-28]:

$$B_0 = 4.0\times10^{-7} T, \quad k = 1.0\times10^{-10} cm^{-1}, \quad T_i = T_{\perp i} = 20 KeV, \quad n_i = n_e = 1 cm^{-3},$$

$$V_A = 8.8\times10^{8} cms^{-1}, \quad v_{T\,e} = 5.9\times10^{9} cms^{-1}, \quad \frac{c_s^2}{V_A^2} = 0.01, \quad \frac{c^2 k_\perp^2}{\omega_{pi}^2} = 0.01.$$

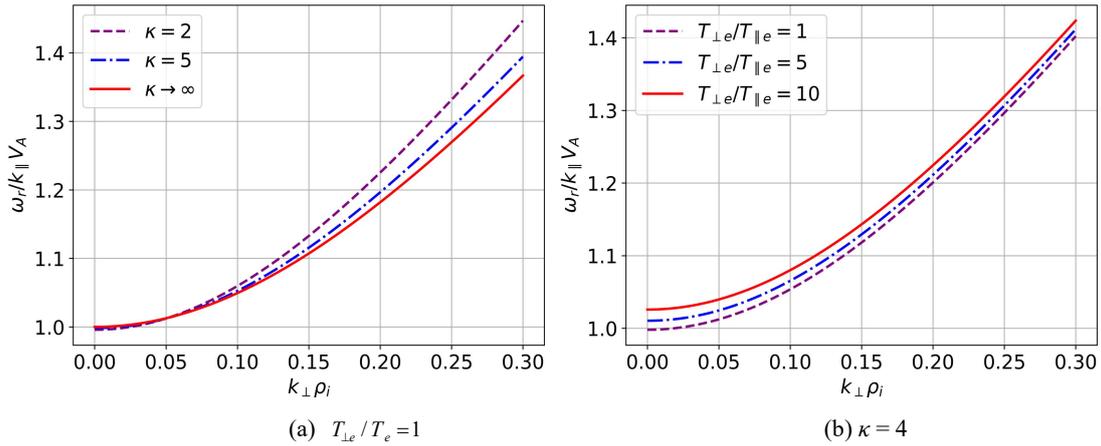

(a) $T_{\perp e}/T_e = 1$  (b) $\kappa = 4$

**Figure 1.** $\omega_r/kV_A$ as a function of perpendicular wavenumber $k_\perp \rho_i$ of kinetic Alfvén waves.

Figs. 1(a)-1(b) show the change of the wave frequency $\omega_r/kV_A$ as a function of perpendicular wavenumber $k_\perp \rho_i$ of kinetic Alfvén waves with the parameter $\kappa$ and electrons temperature anisotropic $T_{\perp e}/T_e$, where Fig. 1(a) is for the parameter $T_{\perp e}/T_e = 1$, and the line with $\kappa \to \infty$ is corresponding to the Maxwellian case of the plasma; Fig. 1(b) is for the parameter $\kappa = 4$.

Fig. 1 shows that with an increase of the perpendicular wavenumber $k_\perp \rho_i$, the wave frequency $\omega_r / k V_A$ of kinetic Alfvén waves will increase monotonically. From Figs. 1(a) we can find that the wave frequency decreases as the $\kappa$-parameter increases. This also proves the existence of superthermal electrons will enhance the wave frequency of kinetic Alfvén waves. Figs. 1(b) shows the effects of $T_{\perp e} / T_e$ on kinetic Alfvén waves. We found that the wave frequency $\omega_r / k V_A$ increases with the increase of $T_{\perp e} / T_e$. The influence of temperature anisotropy on the wave frequency does not seem particularly significant. The reason is because we consider low $\beta$ case, i. e. $\beta \ll 1$ (see equation (25)).

Fig. 2(a)-2(b) depict the effects of the parameter $\kappa$ and electron temperature anisotropic $T_{\perp e} / T_e$ on the wave frequency $\omega_r / k V_A$ of the modified ion acoustic waves in the same case of the kinetic Alfvén waves. We find that the wave frequency $\omega_r / k V_A$ of the modified ion acoustic waves decreases as the perpendicular wavenumber $k_\perp \rho_i$ increases. In Fig. 2(a), we find that the change of the wave frequency $\omega_r / k V_A$ with the parameter $\kappa$ is not monotonous. In the short-wave region, i.e., $k_\perp \rho_i < 0.15$, the wave frequency decreases as $\kappa$ increases while in the long-wave region, the opposite result is observed. Compared with the kinetic Alfvén waves in the same case, the wave frequency of the modified ion-acoustic waves is lower than that of the kinetic Alfvén waves. Fig. 2(b) shows that the larger parameter $T_{\perp e} / T_e$ will lead to the larger wave frequency $\omega_r / k V_A$.

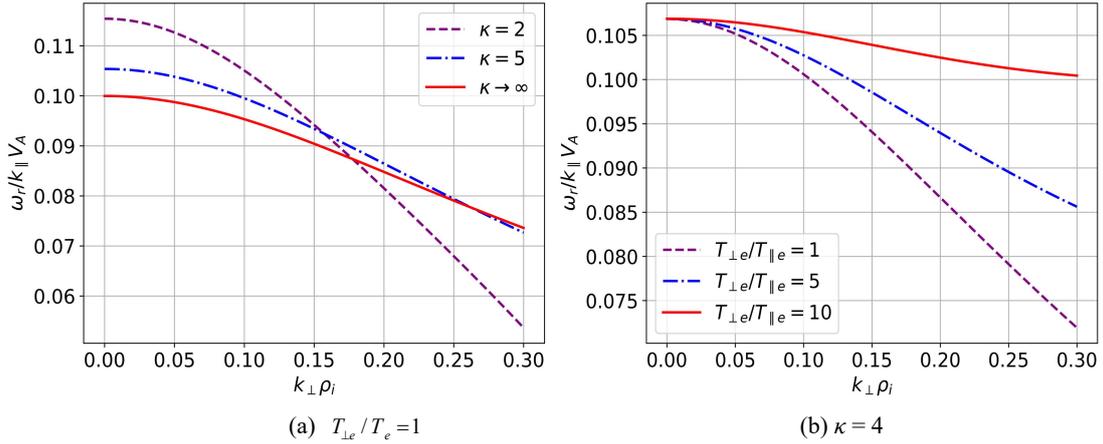

(a) $T_{\perp e} / T_e = 1$  (b) $\kappa = 4$

**Figure 2.** $\omega_r / k V_A$ as a function of perpendicular wavenumber $k_\perp \rho_i$ of modified ion-acoustic waves.

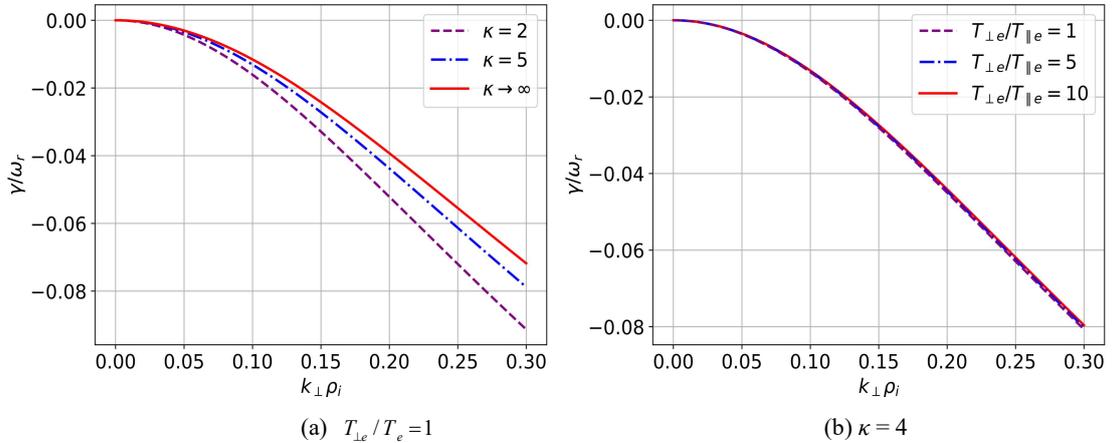

(a) $T_{\perp e} / T_e = 1$  (b) $\kappa = 4$

**Figure 3.** $\gamma / \omega_r$ as a function of perpendicular wavenumber $k_\perp \rho_i$ of kinetic Alfvén waves.

In Fig. 3, we show the variation of the damping rate of the kinetic Alfvén waves as a function of perpendicular wavenumber $k_\perp p_i$ for various values of the parameter $\kappa$ and electrons temperature anisotropic $T_{\perp e}/T_e$, where Fig. 3(a) is for the parameter $T_{\perp e}/T_e = 1$, and the line with $\kappa \to \infty$ is corresponding to the Maxwellian case of the plasma; Fig. 3(b) is for the parameter $\kappa = 4$.

Fig. 3 shows that with the increase of the perpendicular wavenumber $k_\perp p_i$, the damping rate $|\gamma/\omega_r|$ increases monotonically. From Figs. 1(a) we can find that the damping rate in kappa-Maxwellian case is stronger than that in Maxwellian case. The reason may be that the presence of superthermal electrons engages more particles in the resonant wave-particle, leading to an increase in damping rate. Figs. 3(b) shows that the damping rate $|\gamma/\omega_r|$ decreases with the increase of $T_{\perp e}/T_e$.

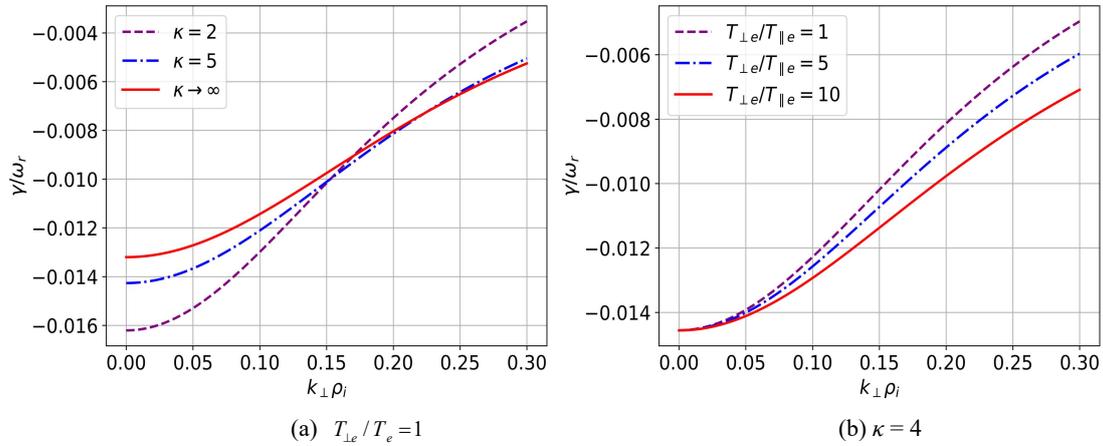

(a) $T_{\perp e}/T_e = 1$  (b) $\kappa = 4$

**Figure 4.** $\gamma/\omega_r$ as a function of perpendicular wavenumber $k_\perp p_i$ of modified ion-acoustic waves.

Figs. 4(a)-(b) describe the change of the damping rate of the modified ion acoustic waves as a function of perpendicular wavenumber for various values of the parameter $\kappa$ and electrons temperature anisotropic $T_{\perp e}/T_e$, where these basic parameters are the same as above. In Fig. 4, it is observed that the damping rate $|\gamma/\omega_r|$ decreases as the perpendicular wavenumber $k_\perp p_i$ increases. From Figs. 4(a) we can find that the damping rate $|\gamma/\omega_r|$ decreases in the short-wave region but increases in the long-wave region as the parameter $\kappa$ increases. Figs. 4(b) shows that the damping rate $|\gamma/\omega_r|$ increases with the increase of the parameter $T_{\perp e}/T_e$.

## 4. Conclusions

In conclusion, we have studied the wave frequency and the damping rate of the kinetic Alfvén waves and the modified ion acoustic waves in temperature anisotropic plasma with a kappa-Maxwellian distribution. We found that both the parameter $\kappa$ and $T_{\perp e}/T_e$ significantly affect the wave frequency and the damping rate of the kinetic Alfvén waves and the modified ion acoustic waves. The wave frequency of kinetic Alfvén waves is larger in kappa-Maxwellian plasma than that in Maxwellian case. The wave frequency of the modified ion-acoustic waves is larger in kappa-Maxwellian plasma than that in Maxwellian case in the short-wave region while in the long-wave region, the opposite result is observed. In the same case, the wave frequency of kinetic Alfvén waves is greater than that of the modified ion acoustic waves. The impact of the parameter $T_{\perp e}/T_e$ on the two modes is relatively small because we consider the low $\beta$ case. Again, we found that the damping rate of kinetic Alfvén waves in kappa-Maxwellian plasma is stronger than that in Maxwellian case. The damping rate of modified ion acoustic waves in kappa-Maxwellian plasma is stronger in the short-wave region but weaker in the long-wave region than that in Maxwellian case. They are also influenced by temperature anisotropic. The damping rate $|\gamma/\omega_r|$ of kinetic Alfvén waves decreases with the increase of the parameter $T_{\perp e}/T_e$ and the damping rate $|\gamma/\omega_r|$ of the modified ion acoustic waves increases with the increase of the parameter $T_{\perp e}/T_e$. These results may help find more interesting applications because kinetic Alfvén waves play an important role in the energy transport of space plasma.

**Appendix A**

For Eq. (11), taking the kappa-Maxwellian distribution $f_{\alpha 0}$ given in Eq. (1) into Eq. (11), using $J_{-n}^2(\mu_\alpha) = J_n^2(\mu_\alpha)$, $d^3v = 2\pi v_\perp dv_\perp dv_\parallel$, we can get that

$$\varepsilon_{xx} = 1 + \sum_\alpha \frac{2\pi \omega_{p\alpha}^2}{\omega} \int v_\perp^2 dv_\perp dv_\parallel \\ \sum_{n=1}^\infty \frac{n^2}{\mu^2} J_n^2(\mu_\alpha) \left( \frac{1}{\omega - k_\parallel v_\parallel - n\Omega_\alpha} + \frac{1}{\omega - k_\parallel v_\parallel + n\Omega_\alpha} \right) \left[ \left(1 - \frac{k_\parallel v_\parallel}{\omega}\right) \frac{\partial f_{\alpha 0}}{\partial v_\perp} + \frac{k_\parallel v_\perp}{\omega} \frac{\partial f_{\alpha 0}}{\partial v_\parallel} \right]. \quad (A1)$$

After some calculations of Eq. (A1), we can get

$$\varepsilon_{xx} = 1 - \sum_\alpha \frac{\omega_{p\alpha}^2}{\omega} \sum_{n=1}^\infty \frac{n^2}{k_\perp^2 \rho_\alpha^2} \times \exp(-k_\perp^2 \rho_\alpha^2) I_n(k_\perp^2 \rho_\alpha^2) \\ \left\{ \frac{\omega}{\omega - n\Omega_\alpha} + \frac{\omega}{\omega + n\Omega_\alpha} + \frac{1}{2}\left(\frac{T_{\parallel \alpha}}{T_{\perp \alpha}} - \frac{\kappa}{\kappa - \frac{3}{2}}\right) \left[ \frac{k_\parallel^2 v_{\parallel \alpha}^2}{(\omega - n\Omega_\alpha)^2} + \frac{k_\parallel^2 v_{\parallel \alpha}^2}{(\omega + n\Omega_\alpha)^2} \right] \right\}, \quad (A2)$$

where $\rho_\alpha^2 = v_{T\perp\alpha}^2 / 2\Omega_\alpha^2$ is the square of the gyroradius of plasma particles. In the same way, we can also get

$$\varepsilon_{zz} = 1 - \sum_\alpha \frac{\omega_{p\alpha}^2}{\omega^2} \exp(-k_\perp^2 \rho_\alpha^2) \left[ I_0(k_\perp^2 \rho_\alpha^2) \eta_\alpha^2 \frac{d}{d\eta_\alpha} Z_{\kappa,M}(\eta_\alpha) - \sum_{n=1}^\infty I_n(k_\perp^2 \rho_\alpha^2) \right] \\ \left[ \omega\left(\frac{1}{\omega - n\Omega_\alpha} + \frac{1}{\omega + n\Omega_\alpha}\right) + n\Omega_\alpha \left( \frac{T_{\parallel \alpha}}{T_{\perp \alpha}} \frac{\kappa \Gamma(\kappa - \frac{3}{2})}{\Gamma(\kappa - \frac{1}{2})} - 1 \right) \left( \frac{1}{\omega - n\Omega_\alpha} - \frac{1}{\omega + n\Omega_\alpha} \right) \right], \quad (A3)$$

where $\eta_\alpha = \omega/(k_\parallel v_{T\parallel\alpha})$, and $Z_{\kappa,M}(\eta_\alpha)$ is the modified plasma dispersion function with the kappa-Maxwellian distribution, given by [18]

$$Z_{\kappa,M}(\eta_\alpha) = \frac{1}{\sqrt{\pi}} \frac{\Gamma(\kappa+1)}{\kappa^{3/2} \Gamma(\kappa-1/2)} \int_{-\infty}^\infty \frac{(1+\kappa^{-1}\xi^2)^{-\kappa}}{\xi - \eta_\alpha} d\xi. \quad (A4)$$

We let $\lambda_\alpha = k_\perp^2 \rho_\alpha^2$ and $A_0(\lambda_\alpha) = \exp(-\lambda_\alpha) I_n(\lambda_\alpha)$ with the first kind modified Bessel function $I_n(z)$, defined by [29]

$$I_n(z) = \frac{(z/2)^n}{\sqrt{\pi}\Gamma(n+\frac{1}{2})} \int_{-1}^1 \exp(-zt)(1-t^2)^{n-1/2} dt,$$

and then Eq. (A2) and Eq. (A3) can be simplified as

$$\varepsilon_{xx} = 1 - \sum_\alpha \frac{\omega_{p\alpha}^2}{\omega^2} \sum_{n=1}^\infty \frac{n^2}{k_\perp^2 \rho_\alpha^2} A_n(\lambda_\alpha) \left[ \frac{2\omega^2}{\omega^2 - n^2\Omega_\alpha^2} + k_\parallel^2 v_{\parallel\alpha}^2 \left( \frac{T_{\perp\alpha}}{T_{\parallel\alpha}} - \frac{\kappa}{\kappa - \frac{3}{2}} \right) \left( \frac{\omega^2 + n^2\Omega_\alpha^2}{(\omega^2 - n^2\Omega_\alpha^2)^2} \right) \right], \quad (A5)$$

$$\varepsilon_{zz} = 1 - \sum_\alpha \frac{\omega_{p\alpha}^2}{\omega^2} A_0(\lambda_\alpha) \eta_\alpha^2 \frac{d}{d\eta_\alpha} Z_{\kappa,M}(\eta_\alpha)$$
$$- \sum_\alpha \frac{2\omega_{p\alpha}^2}{\omega^2} \sum_{n=1}^\infty A_n(\lambda_\alpha) \left[ 1 + \frac{T_{\|\alpha}}{T_{\perp\alpha}} \frac{\kappa \Gamma(\kappa - \frac{3}{2})}{\Gamma(\kappa - \frac{1}{2})} \frac{n^2 \Omega_\alpha^2}{\omega^2 - n^2 \Omega_\alpha^2} \right]. \tag{A6}$$

By using low-frequency limit, i.e., $\omega \ll n\Omega_\alpha$, and the identity of the modified Bessel function, i.e., $2\sum_{n=1}^\infty A_n(\lambda_\alpha) = 1 - A_0(\lambda_\alpha)$, they can be further simplified as

$$\varepsilon_{xx} = 1 + \sum_\alpha \frac{\omega_{p\alpha}^2}{\Omega_\alpha^2} \left( \frac{1 - A_0(\lambda_\alpha)}{\lambda_\alpha} \right) \left[ 1 - \frac{k_\|^2 v_{\|\alpha}^2}{2\omega^2} \left( \frac{T_{\perp\alpha}}{T_{\|\alpha}} - \frac{\kappa}{\kappa - \frac{3}{2}} \right) \right], \tag{A7}$$

$$\varepsilon_{zz} = 1 - \sum_\alpha \frac{\omega_{p\alpha}^2}{\omega^2} A_0(\lambda_\alpha) \eta_\alpha^2 \frac{d}{d\eta_\alpha} Z_{\kappa,M}(\eta_\alpha) + \sum_\alpha \frac{\omega_{p\alpha}^2}{\omega^2} \left[ 1 - A_0(\lambda_\alpha) \right] \left( \frac{T_{\|\alpha}}{T_{\perp\alpha}} \frac{\kappa \Gamma(\kappa - \frac{3}{2})}{\Gamma(\kappa - \frac{1}{2})} - 1 \right). \tag{A8}$$

For compensated electron-ion plasma, $\varepsilon_{xx}$ and $\varepsilon_{zz}$ can be rewritten as

$$\varepsilon_{xx} = 1 + \frac{c^2}{V_A^2} \left( \frac{1 - A_0(\lambda_i)}{\lambda_i} \right) - \frac{c^2 k_\|^2}{\omega^2} \frac{c_s^2}{V_A^2} \left[ \left( \frac{1 - A_0(\lambda_e)}{\lambda_e} \right) \left( \frac{\kappa - \frac{3}{2}}{\kappa} \frac{T_{\perp e}}{T_{\|e}} - 1 \right) + \left( \frac{1 - A_0(\lambda_i)}{\lambda_i} \right) \left( \frac{\kappa - \frac{3}{2}}{\kappa} \frac{T_{\perp i}}{T_{\|i}} - 1 \right) \frac{T_{\|i}}{T_{\|e}} \right]$$

$$= 1 + \frac{c^2}{V_A^2} \left( \frac{1 - A_0(\lambda_i)}{\lambda_i} \right) - \frac{c^2 k_\|^2}{\omega^2} \chi_1, \tag{A9}$$

$$\varepsilon_{zz} = 1 - \frac{\omega_{pe}^2}{k_\|^2 v_{T\|e}^2} A_0(\lambda_e) \frac{d}{d\eta_e} Z_{\kappa,M}(\eta_e) - \frac{\omega_{pi}^2}{k_\|^2 v_{T\|i}^2} A_0(\lambda_i) \frac{d}{d\eta_i} Z_{\kappa,M}(\eta_i)$$
$$+ \frac{\omega_{pe}^2}{\omega^2} (1 - A_0(\lambda_e)) \left[ \frac{T_{\|e}}{T_{\perp e}} \frac{\kappa \Gamma(\kappa - \frac{3}{2})}{\Gamma(\kappa - \frac{1}{2})} - 1 \right] + \frac{\omega_{pi}^2}{\omega^2} (1 - A_0(\lambda_{ei})) \left[ \frac{T_{\|i}}{T_{\perp i}} \frac{\kappa \Gamma(\kappa - \frac{3}{2})}{\Gamma(\kappa - \frac{1}{2})} - 1 \right]$$

$$= 1 - \frac{A_0(\lambda_e)}{2 k_\|^2 \lambda_{De}^2} \frac{d}{d\eta_e} Z_{\kappa,M}(\eta_e) - \frac{A_0(\lambda_i)}{2 k_\|^2 \lambda_{Di}^2} \frac{d}{d\eta_i} Z_{\kappa,M}(\eta_i) + \frac{\omega_{pi}^2}{\omega^2} \chi_2, \tag{A10}$$

where $c_s = \sqrt{T_{\|e}/m_i}$, $V_A = B_0 / \sqrt{4\pi n_0 m_i}$ and $\lambda_{D\alpha}^2 = v_{T\|\alpha}^2 / 2\omega_{p\alpha}^2$.

**Acknowledgments**

Ran Guo is supported by the National Natural Science Foundation of China under Grant No. 12105361.